\documentclass[9pt,twocolumn,twoside]{opticajnl}
\journal{opticajournal} 

\setboolean{shortarticle}{true}

\usepackage{lineno}

\title{Calibrated Plane--Convex Microcavity for Room-Temperature Polaritons with Geometric \(g\)-Scaling}
\author{Ling-Qi Huang}
\author{Shih-Chung Chen}
\author{Chia-Hao Lin}
\author{Li-Tzu Wang}
\author{Khemendra Shukla}
\author{Kai-Peng Hsieh}
\author{Leng-Hsien Huang}
\author[*]{Tsung-Sheng Kao}
\author[*]{Hyeyoung Ahn}
\author[*]{Tzu-Ling Chen}

\affil{Department of Photonics, College of Electrical and Computer Engineering, National Yang Ming Chiao Tung University, Hsinchu, 30010, Taiwan}

\affil[*]{tlc@nycu.edu.tw}

\begin{abstract}
We present a plane--convex open microcavity that enables room-temperature polariton spectroscopy with a simple geometric handle on the coupling rate. The effective length \(L_{\mathrm{eff}}\) is absolutely calibrated from the free-spectral range, and piezo tuning is performed at near-normal incidence (\(k_{\parallel}\!\approx\!0\)) to avoid angle-induced degradation.
Using spin-coated PEA\(_2\)PbI\(_4\) quasi-2D perovskites, we observe clear anti-crossings in reflection, with vacuum Rabi splittings up to \(79~\mathrm{meV}\) (reflection) and \(84~\mathrm{meV}\) (PL) near \(L_{\mathrm{eff}}\!\approx\!3.4~\mu\mathrm{m}\).
A linewidth-corrected analysis converts the apparent splitting into the coherent exciton-photon coupling rate \(g\), revealing a robust geometric scaling \(g \propto L_{\mathrm{eff}}^{-1/2}\) across multiple longitudinal orders and spatial sites, consistent with the filled-mode thin-film limit where transverse area cancels in the mode volume.
The platform establishes a compact, broadly compatible testbed for room-temperature polaritons and provides a practical design rule: shortening \(L_{\mathrm{eff}}\) is a reliable, geometric lever to strengthen collective coupling in plane--convex microcavities.
\end{abstract}

\setboolean{displaycopyright}{false} 

\begin{document}

\maketitle


Strong light--matter coupling in optical microcavities hybridizes cavity photons and excitons into upper/lower polaritons with a resolvable Rabi splitting, enabling room-temperature condensation, nonlinear optics, and integrated polaritonic functions \cite{lidzey1999room,kasprzak2006bose,gu2021enhanced,chikkaraddy2016single}, and motivating emerging studies of cavity-QED–mediated chemical reactivity \cite{thomas2019tilting,hsu2025chemistry}. Achieving large, clearly resolved splittings at ambient conditions requires maximizing the coherent exciton-photon coupling rate $g$ while controlling dissipation. Two-dimensional and layered semiconductors are ideal emitters thanks to their large oscillator strengths and robust excitons \cite{lidzey1998strong,wang2016coherent}.

Within a coupled-oscillator description, the near–zero-detuning splitting satisfies \(\hbar\Omega \simeq 2g\) when cavity and exciton dephasing rates are small \((\kappa,\gamma_e \ll g)\) and comparable \cite{savona1995quantum,sanchez2022theoretical}. Thus, enlarging the vacuum Rabi splitting amounts to increasing \(g\).
The coupling rate scales inversely with the square root of the effective electromagnetic mode volume ($V_{\mathrm{eff}}$),
\begin{equation}
g \;\propto\; \frac{1}{\sqrt{V_{\mathrm{eff}}}},\qquad
V_{\mathrm{eff}} \;\propto\; \pi w_0^2 L_{\mathrm{eff}}.
\end{equation}
For a plane–concave cavity, the fundamental Gaussian mode follows ABCD optics with a well-defined waist ($w_0$); at the plane mirror \(w_0^2 = \frac{\lambda}{\pi}\sqrt{L_{\mathrm{eff}}\!\left(R-L_{\mathrm{eff}}\right)}\).
In the common limit \(L_{\mathrm{eff}}\!\ll\!R\) (concave mirror ROC), this gives \(w_0^2 \propto \sqrt{L_{\mathrm{eff}}}\).
By contrast, in near-instability geometries such as plane–plane or plane–convex, no closed-form stable Gaussian eigenmode (and hence no unique \(V_{\mathrm{eff}}\)) exists. The dependence \(g(L_{\mathrm{eff}})\) must be established empirically.

Reducing the cavity volume, most effectively by shortening \(L_{\mathrm{eff}}\) while keeping \(w_0\) stable, is therefore the primary lever for stronger
coupling.
Fixed-length DBR stacks and monolithic Fabry–Pérot (FP) resonators have demonstrated efficient light-matter coupling \cite{hirai2023molecular,timur2020mechanisms,chen20242d,liu2015strong}, but they offer limited in situ control and cumbersome sample exchange. Angle-resolved tuning can also degrade finesse at large incidence angles. Open-access microcavities overcome these constraints by varying the physical length with piezo actuators while preserving free-space access \cite{li2019tunable,flatten2016room,hirai2024optical,konrad2015controlling}. However, broadly applicable procedures that keep finesse high during tuning, provide traceable, absolute calibration of the effective optical length $L_{\mathrm{eff}}$, and enable quantitative comparisons across materials and varies cavity longitudinal mode number remain scarce.


Here we realize a finesse-preserving, open-access \emph{plane–convex} microcavity that provides continuous length tuning down to \(L_{\mathrm{eff}}\!\approx\!1.2~\mu\mathrm{m}\) (DBR-limited), absolutely calibrated via the Free-Spectral Range (FSR), while maintaining \(k_{\parallel}\!\approx\!0\) throughout the sweep. Using spin-coated PEA\(_2\)PbI\(_4\) quasi-2D perovskites, we observe clear room-temperature anti-crossings and a vacuum Rabi splitting of \(79~\mathrm{meV}\) (reflection) at \(L_{\mathrm{eff}}\!\approx\!3.4~\mu\mathrm{m}\). Measurements across multiple longitudinal orders reveal a simple geometric scaling, \(g \propto L_{\mathrm{eff}}^{-1/2}\), consistent with a filled-mode thin-film picture in which the lateral dipole number cancels the transverse mode area.


\begin{figure*}[t]
    \centering
    \includegraphics[width=0.9\linewidth]{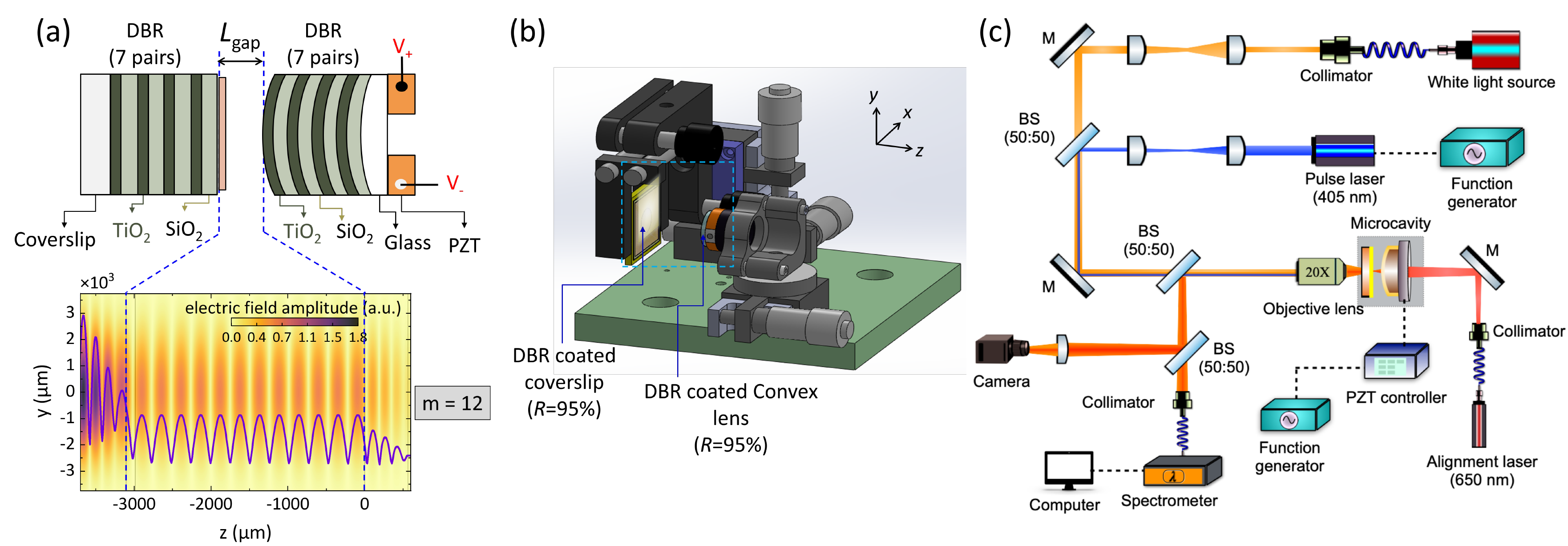}
    \caption{\textbf{Experimental setup and plane–convex microcavity.}
    \textbf{(a)} Cross-section of the cavity: a SiO$_2$-terminated DBR on a borosilicate coverslip hosts the sample; the opposing mirror is a TiO$_2$-terminated DBR on a convex substrate bonded to a ring PZT for axial actuation. (Bottom inset) transfer-matrix field profile showing an antinode at the planar DBR surface.
    \textbf{(b)} Opto-mechanical assembly providing independent lateral (XY) translation and tip–tilt control for both mirrors, with the ring PZT setting $L_{\mathrm{eff}}$.
\textbf{(c)} Optical layout for white-light reflectance and PL spectroscopy, assisted by a 650\,nm alignment beam. A $20\times$ objective lens mode-matches the input beam to the cavity, while a spectrometer and camera record spectra and images, respectively.}
    \label{fig:setup_mechanics}
\end{figure*}

The plane--convex open-access micro-cavity in Fig.~\ref{fig:setup_mechanics}\textbf{(a)} comprises a 0.17~mm borosilicate coverslip (planar mirror) and a convex mirror (1/2" diameter, radius of curvature 51.5~mm), both coated with high-reflectivity DBRs fabricated at the HOPE Laboratory, Photonics Research Center, National Tsing Hua University. Each mirror employs 7.5-pair SiO$_2$/TiO$_2$ quarter-wave pairs, yielding a reflectivity of $95\pm0.7\%$ across the wavelength range of 520--650~nm and defining a tunable air gap $L_{\mathrm{gap}}$.
The planar DBR is terminated with a low-index SiO$_2$ layer to maximize the electric field at the sample position and enhance light--matter coupling, while the opposing covex mirror is terminated with a TiO$_2$ layer.

The cavity resonance is set by the longitudinal condition,
\begin{equation}
m\,\frac{\lambda_c}{2} \;=\; n_{\mathrm{eff}}\,L_{\mathrm{eff}},
\label{eq:res_condition}
\end{equation}
where \(m\in\mathbb{Z}\) is the longitudinal mode index. The index \(m\) fixes the standing-wave order, and therefore the axial pattern of nodes and antinodes of the intracavity field. This directly governs exciton–field overlap: positioning a thin emitter at an antinode (and matching its thickness to the field period) maximizes dipole–mode coupling, whereas off-antinode placement reduces the effective interaction \cite{ebbesen2016hybrid}. As shown in Fig.~\ref{fig:setup_mechanics}\textbf{(a)} (bottom), the simulated axial field profile peaks at the planar mirror, where the active layer is located.

Figure~\ref{fig:setup_mechanics}\textbf{(b)} shows the mechanical design of the microcavity, where the convex DBR mirror was mount on a two-axis adjustable holder (POLARIS-K05S1, $\pm$5$^{\circ}$ angular range, 11~mrad/rev), which provide the primary angular control required for preserving mirror parallelism and ensure optimized cavity alignment. Coarse cavity-length adjustment was achieved using an one-axis translation stage with a 3~mm traveling range, following by fine tuning with a piezoelectric actuator.
For precise spatial positioning of the sample, the planar DBR (sample side) was mounted on a two-adjuster tilt mount combined with a two-axis translation stage for lateral alignment.


The overall optical layout is shown in Fig.~\ref{fig:setup_mechanics}\textbf{(c)}, a broadband tungsten--halogen lamp (SLS201L, Thorlabs) serves as the white-light source for reflectance spectroscopy (probing the optical states), whereas a 405~nm externally modulated pulses laser (pulse width $\approx$ 177~$\mu\text{s}$, repetition rate $\approx$ 692~Hz) serves as the excitation source for photoluminescence (PL), probing the excited-state population.
To achieve optimal mode matching and minimize the spot size on the sample (white-light spot $\approx$ 15~$\mu\text{m}$, 405~nm excitation laser spot $\approx$ 5~$\mu\text{m}$), a 20$\times$ objective lens was used for focusing.

\begin{figure}[t]
    \centering
    \includegraphics[width=\linewidth]{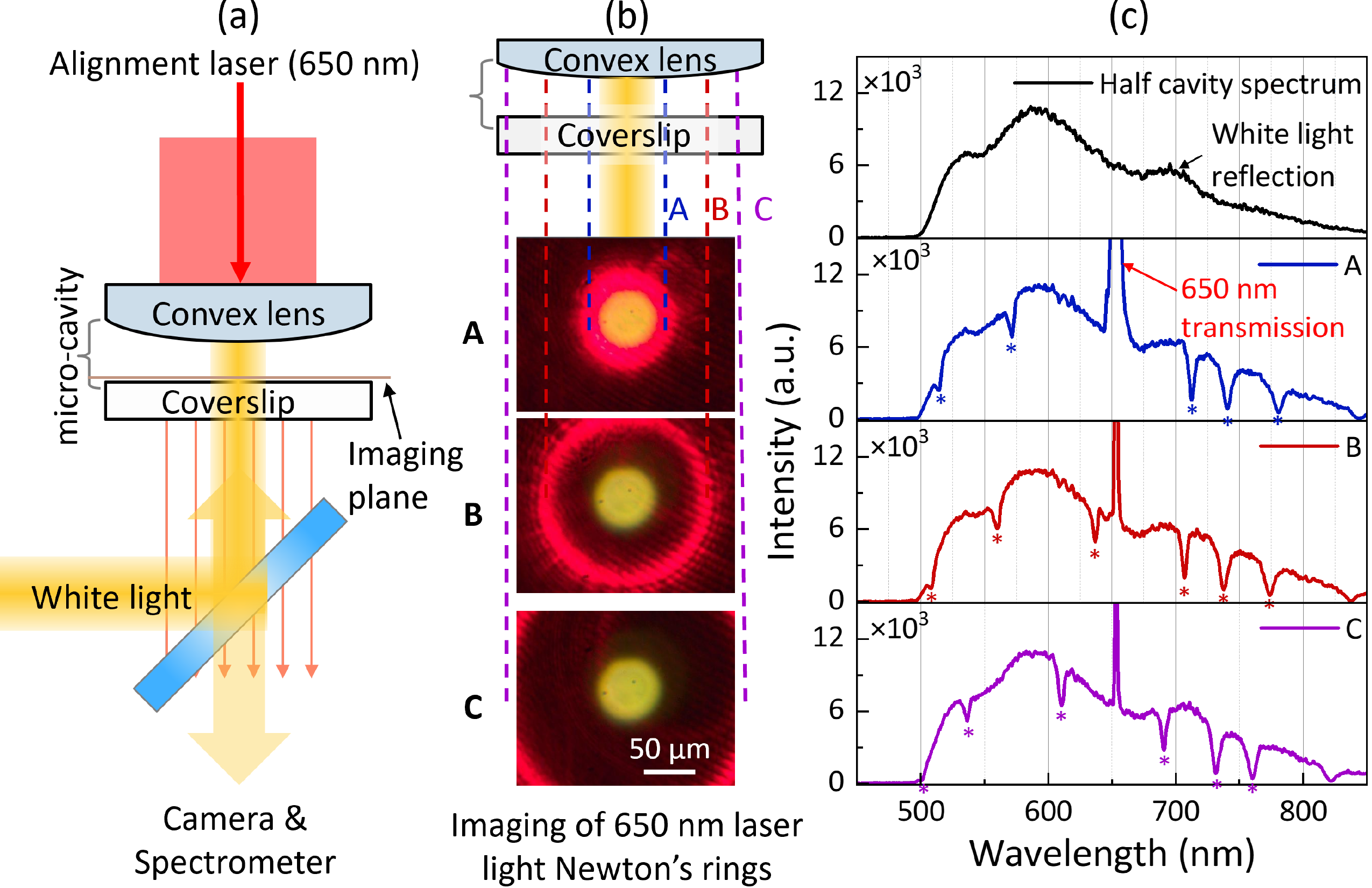}
    \caption{\textbf{Alignment workflow.}
\textbf{(a)} Concept: co-monitor white-light reflection and 650\,nm transmission while adjusting tip–tilt and axial position of the convex lens. 
\textbf{(b)} Camera frames at three lateral sites (A, B, C) with the cavity length tuned to three distinct values. The Newton-ring radius varies with the local gap, and 650\,nm transmission appears only at resonance. The corresponding co-recorded spectra at A, B, and C are shown in \textbf{(c)}, where the longitudinal modes appear as reflection dips. At resonance a narrow 650\,nm transmission peak is observed.}
    \label{fig:alignment_procedure}
\end{figure}

The signal reflected from the micro-cavity on the sample side is directed to a beam splitter, where it is divided into two paths: one directed to a high-resolution spectrometer (Andor Kymera 328i spectrograph; spectral resolution $\approx$ 0.31--0.44~nm) via a fiber-coupled setup for spectral acquisition. During a single PZT scan, each spectrum was acquired with an exposure time of 200~ms.

\begin{figure}[t]
    \centering
    \includegraphics[width=0.75\linewidth]{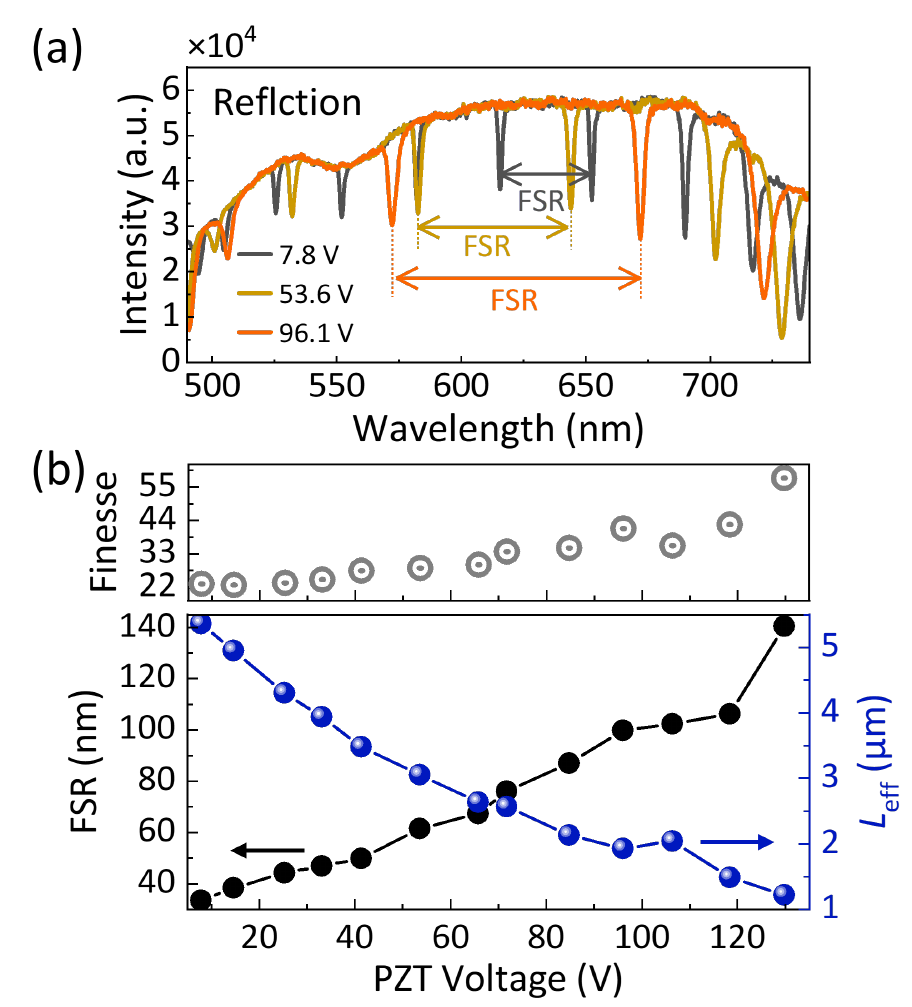}
    \caption{\textbf{Plano--convex microcavity and FSR calibration.}
    \textbf{(a)} Bare-cavity reflection spectra at several PZT voltages; dip spacings yield the FSR.
    \textbf{(b)} FSR (black) and derived $L_{\mathrm{eff}}$ (blue) vs.\ PZT bias, demonstrating continuous tuning down to $\sim$1.2~$\mu\mathrm{m}$.}
    \label{fig:plano_convex_fsr}
\end{figure}

To optimize light–cavity coupling and minimize the cavity length, we combine white-light reflection with 650~nm transmission imaging to center the beam on the convex mirror apex [Fig.~\ref{fig:alignment_procedure}]. The 650~nm beam produces transmitted Newton’s rings that provide tip–tilt feedback, locate the minimum-gap region, and verify mirror parallelism. When the mirrors approach, a clear interference pattern appears on the camera [Fig.~\ref{fig:alignment_procedure}\textbf{(b)}], with co-recorded reflection spectra shown in [Fig.~\ref{fig:alignment_procedure}\textbf{(c)}]. Fine tilt adjustments align the ring center with the detection area, yielding the strongest, most stable cavity resonance.

Figure~\ref{fig:plano_convex_fsr}\textbf{(a)} shows bare-cavity reflection spectra acquired at different PZT voltages. From these spectra we extract the FSR and use it to derive the effective cavity length \(L_{\mathrm{eff}}\).
At normal incidence the frequency-domain FSR is
\begin{equation}
\mathrm{FSR}_{\nu}=\frac{c}{2\,n_{\mathrm{eff}}\,L_{\mathrm{eff}}},
\label{eq:fsr_freq}
\end{equation}
with \(c\) the speed of light and \(n_{\mathrm{eff}}\!\approx\!1\) for an air-gap cavity. Equivalently, near a central wavelength \(\lambda_c\),
\(\mathrm{FSR}_{\lambda}\!\approx\!\lambda_c^{2}/(2\,n_{\mathrm{eff}}\,L_{\mathrm{eff}})\).
The shortest attainable length in the empty cavity is
\(L_{\mathrm{eff}}\!\approx\!1.2~\mu\text{m}\), inferred from the largest measured wavelength-domain FSR of \(\sim\!140~\text{nm}\)
[Fig.~\ref{fig:plano_convex_fsr}\,\textbf{(b)}].
This is primarily limited by the DBR penetration depths on the two mirrors; for the terminating layers we estimate
\(L_{\mathrm{pen}}\!\approx\!0.15~\mu\text{m}\) (TiO\(_2\)) and \(\approx\!0.55~\mu\text{m}\) (SiO\(_2\)) \cite{koks2021microcavity}.
Once \(L_{\mathrm{eff}}(V_{\mathrm{PZT}})\) is determined, the nearest longitudinal mode index \(m\) is assigned via \eqref{eq:res_condition}.

\begin{figure}[t]
    \centering
    \includegraphics[width=1\linewidth]{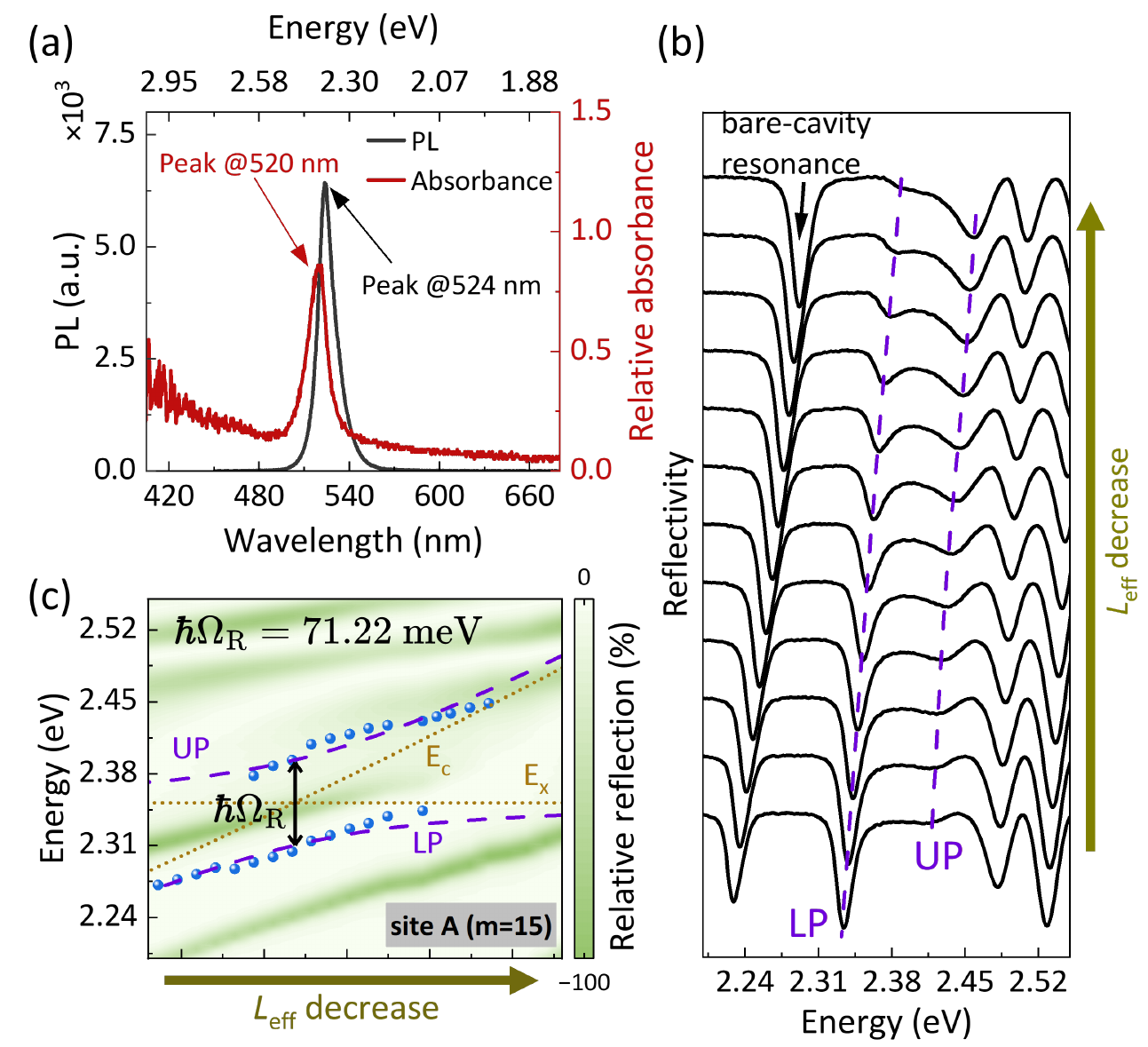}
    \caption{\textbf{Energy-resolved reflectivity of the tunable microcavity.}
    \textbf{(a)} PEA$_2$PbI$_4$ film on a planar DBR: excitonic absorption ($\sim$520\,nm) and resonant PL ($\sim$524\,nm). 
    \textbf{(b)} Stacked reflection spectra while $L_{\mathrm{eff}}$ is scanned. A bare-cavity resonance in the non-absorbing region keeps a constant linewidth, indicating finesse is preserved during PZT tuning. Near $E_x$, the resonance splits into UP/LP. 
    \textbf{(c)} Reflection-intensity map versus wavelength and $L_{\mathrm{eff}}$ with fitted UP/LP (blue) and coupled-oscillator branches (purple). The minimum splitting at (near) zero detuning is $\hbar\Omega_R \approx 71$\,meV.}
    \label{fig:rabi}
\end{figure}

The PEA$_2$PbI$_4$ films on the planar DBR exhibit a pronounced excitonic absorption at $\lambda_{\mathrm{abs}}\!\approx\!520$~nm and resonant photoluminescence at $\lambda_{\mathrm{PL}}\!\approx\!524$~nm [Fig.~\ref{fig:rabi}\textbf{(a)}]. The narrow spectral features are consistent with strong excitonic confinement characteristic of quasi-2D perovskites (here, nominal \(n=1\)).
The perovskite precursor was prepared by dissolving 51.9~mg PEAI and 48.1~mg PbI$_2$ in 1000~$\mu\text{L}$ DMF, followed by stirring at $65~^\circ$C for 30~min. During stirring, DBR substrates were treated by UV–ozone for 30~min to remove residual organics and increase surface hydrophilicity. A 50~$\mu\text{L}$ aliquot of the precursor was dispensed onto the DBR and spin-coated at 6000~rpm for 20~s, then annealed at $100~^\circ$C for 1~min.

\begin{figure}[t]
    \centering
    \includegraphics[width=\linewidth]{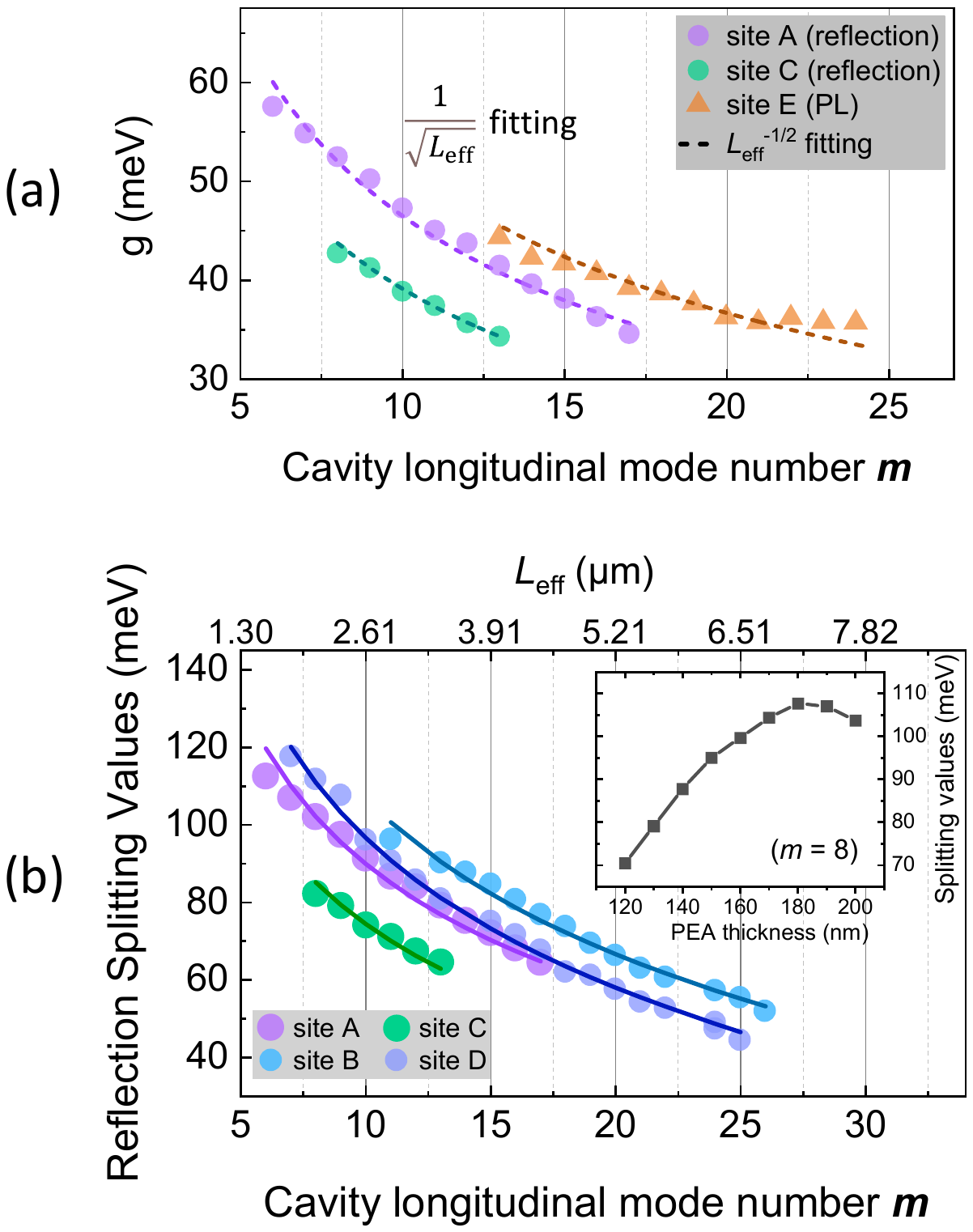}
\caption{\textbf{Splitting and coupling vs.\ cavity length.}
\textbf{(a)} Coupling rate \(g\) from Eq.~(\ref{eq:g_from_split}) vs.\ \(L_{\mathrm{eff}}\) for reflection (circles) and PL (triangles). Dashed curves are \(L_{\mathrm{eff}}^{-1/2}\) fits.
\textbf{(b)} Reflection-derived Rabi splittings vs.\ longitudinal order \(m\). The solid line is a fit using the Savona analytic model with \(g(L_{\mathrm{eff}})\propto L_{\mathrm{eff}}^{-1/2}\). Inset: FDTD simulated thickness dependence of the vacuum Rabi splitting at fixed cavity order.}
    \label{fig:g-length}
\end{figure}

Sweeping the cavity length with the PZT tunes the bare-cavity mode \(E_c = hc/\lambda_c\) through the exciton energy \(E_x\), yielding a clear anti-crossing in reflection [Fig.~\ref{fig:rabi}\textbf{(b,c)}] and PL [Fig. S1]. Upper- and lower-polariton (UP/LP) peak centers are obtained by Lorentzian fits (fitting uncertainty \(<\!5\%\)) and compared with the coupled-oscillator eigenvalues \cite{canales2023polaritonic}:
\begin{equation}
E_{\pm}=\frac{E_c+E_x}{2}\;\pm\;\frac{1}{2}\sqrt{\Delta^2+4g^2},
\qquad \Delta \equiv E_c-E_x .
\label{eq:co_eigs}
\end{equation}
At (near) zero detuning (\(\Delta \approx 0\)), the fitted normal-mode separation gives \(\hbar\Omega_R \approx 71~\mathrm{meV}\) (site~A in Fig.~\ref{fig:g-length}). 

To incorporate finite linewidths, we convert the apparent splitting into a
coupling rate via
\begin{equation}
g \;=\; \sqrt{\left(\frac{\hbar\Omega_R}{2}\right)^2
          + \left(\frac{\kappa-\gamma}{4}\right)^2},
\label{eq:g_from_split}
\end{equation}
where the dephasing rates \(\kappa\) and \(\gamma\) are measured using the HWHM energy linewidths of the bare cavity mode and exciton, respectively. The exciton linewidth \(\gamma\) is taken from single-pass absorption/PL of the perovskite film, and the cavity linewidth \(\kappa\) from the bare-cavity reflection dip. Across our datasets, we find \(\gamma \approx 26\!-\!36~\mathrm{meV}\) and
\(\kappa \approx 7\!-\!11~\mathrm{meV}\). 

Figure~\ref{fig:g-length}\textbf{(a)} plots the linewidth-corrected \(g\) versus the longitudinal order \(m\), with colors indicating distinct measurement sites on the same film. The mode index \(m\) is assigned from the calibrated bare-cavity FSR and decreases consecutively as the cavity is scanned in length (details in Fig.~S2).
Coupling strengths extracted at three representative sites (from both reflection- and PL-derived splittings) collapse onto a single geometric law,
\[
g \;\propto\; L_{\mathrm{eff}}^{-1/2},
\]
as shown in Fig.~\ref{fig:g-length}\textbf{(a)}. 
This scaling follows the filled–mode thin–film picture: when the perovskite laterally covers the optical mode, \(N \!\propto\! \pi w_0^2 t\) while \(V_{\mathrm{mode}} \!\propto\! \pi w_0^2 L_{\mathrm{eff}}\); the transverse area cancels, giving \(g \propto \sqrt{t/L_{\mathrm{eff}}}\). The large mirror radius in our plane–convex geometry keeps \(w_0\) nearly constant over the tuning range, enhancing this scaling.

Using the empirical \(g(L_{\mathrm{eff}})\propto L_{\mathrm{eff}}^{-1/2}\), we fit the measured reflection splittings versus \(m\) with the Savona analytic expression for the observable zero-detuning splitting [Eq.~(19) in Ref.~\cite{savona1995quantum}. See Sec.~S4 of the SI for details].
The solid curve in Fig.~\ref{fig:g-length}(b) captures the overall trend and site-to-site spread.
The site-to-site variation on a single sample is attributed to film-thickness nonuniformity. Three-dimensional finite-difference time-domain (FDTD) simulations quantify the sensitivity of the Rabi splitting to the perovskite thickness and reproduce the observed scatter across sites. The measured thickness varies from (\(\sim 135\!-\!160\)~nm), and Fig.~S3 shows that, over this range, the splitting depends approximately linearly on thickness..


In conclusion, we have realized a finesse-preserving, open-access plane–convex microcavity with FSR-calibrated length control down to \(L_{\mathrm{eff}}\!\approx\!1.2~\mu\mathrm{m}\),
providing a compact platform for quantitative polariton studies. Using spin-coated PEA\(_2\)PbI\(_4\) at room temperature, we observed clear anti-crossings in reflection/PL and extracted vacuum Rabi splittings up to \(79\)~meV (reflection) and \(84\)~meV (PL) near \(L_{\mathrm{eff}}\!\approx\!3.4~\mu\mathrm{m}\).
After correcting for finite linewidths, the coupling rate follows a simple geometric law, \(g \propto L_{\mathrm{eff}}^{-1/2}\), across multiple longitudinal orders and spatial sites.

Even in the absence of a closed‐form Gaussian mode and a unique \(V_{\mathrm{eff}}\) for plane–convex geometries, the measured \(g(L_{\mathrm{eff}})\) establishes a practical and portable rule: shortening \(L_{\mathrm{eff}}\) reliably strengthens the collective coupling (\(g \propto L_{\mathrm{eff}}^{-1/2}\)). This provides a clear design guideline for open plane–convex microcavities.

Because the platform separates length tuning from angular dispersion, supports rapid sample exchange, and enables traceable comparisons between reflection and PL, it constitutes a versatile testbed for room‐temperature polaritons. The approach is readily extendable to other thin‐film emitters, time‐resolved measurements, electrically assisted tuning, and device‐oriented polaritonic architectures.


\begin{backmatter}
\bmsection{Funding} This work was supported by the National Science and Technology Council (NSTC), Taiwan, under grants NSTC 113-2628-M-A49-002, and by the Ministry of Education (Taiwan) through the Yushan Young Scholar Program. H.-Y.~A. acknowledges support from NSTC 113-2112-M-A49-021-MY3. T.-S.~Kao acknowledges support from NSTC 113-2221-E-A49-053. L.-Q. Huang thanks the support from NSTC Graduate Research Fellowship (NSTC-GRF).

\bmsection{Acknowledgment} 
T.-L.~Chen thanks Prof.~Jhih-Sheng Wu for valuable discussions.
\bmsection{Disclosures} The authors declare no conflict of interest.

\smallskip

\bmsection{Data Availability Statement} Data underlying the results presented in this paper are not publicly available at this time but can be obtained from the authors upon reasonable request.


\end{backmatter}


\bibliography{sample}

@article{flatten2016room,
  title={Room-temperature exciton-polaritons with two-dimensional WS2},
  author={Flatten, Lucas C and He, Zhengyu and Coles, David M and Trichet, Aurelien AP and Powell, Alex W and Taylor, Robert A and Warner, Jamie H and Smith, Jason M},
  journal={Scientific reports},
  volume={6},
  number={1},
  pages={33134},
  year={2016},
  publisher={Nature Publishing Group UK London}
}

@article{timur2020mechanisms,
  title={Mechanisms of blueshifts in organic polariton condensates},
  author={Timur, Yagafarov and Denis, Sannikov and Anton, Zasedatelev and Kyriacos, Georgiou and Anton, Baranikov and Oleksandr, Kyriienko and Ivan, Shelykh and Lizhi, Gai and Shen, Zhen and Lidzey, David and others},
  journal={Communications Physics},
  volume={3},
  number={1},
  year={2020},
  publisher={Nature Publishing Group}
}

@article{sanchez2022theoretical,
  title={A theoretical perspective on molecular polaritonics},
  author={S{\'a}nchez-Barquilla, M{\'o}nica and Fern{\'a}ndez-Dom{\'\i}nguez, Antonio I and Feist, Johannes and Garc{\'\i}a-Vidal, Francisco J},
  journal={ACS photonics},
  volume={9},
  number={6},
  pages={1830--1841},
  year={2022},
  publisher={ACS Publications}
}

@article{hsu2025chemistry,
  title={Chemistry Meets Plasmon Polaritons and Cavity Photons: A Perspective from Macroscopic Quantum Electrodynamics},
  author={Hsu, Liang-Yan},
  journal={The Journal of Physical Chemistry Letters},
  volume={16},
  number={6},
  pages={1604--1619},
  year={2025},
  publisher={ACS Publications}
}

@article{thomas2019tilting,
  title={Tilting a ground-state reactivity landscape by vibrational strong coupling},
  author={Thomas, Anoop and Lethuillier-Karl, Lucas and Nagarajan, Kalaivanan and Vergauwe, Robrecht MA and George, Jino and Chervy, Thibault and Shalabney, Atef and Devaux, Elo{\"\i}se and Genet, Cyriaque and Moran, Joseph and others},
  journal={Science},
  volume={363},
  number={6427},
  pages={615--619},
  year={2019},
  publisher={American Association for the Advancement of Science}
}

@article{chikkaraddy2016single,
  title={Single-molecule strong coupling at room temperature in plasmonic nanocavities},
  author={Chikkaraddy, Rohit and De Nijs, Bart and Benz, Felix and Barrow, Steven J and Scherman, Oren A and Rosta, Edina and Demetriadou, Angela and Fox, Peter and Hess, Ortwin and Baumberg, Jeremy J},
  journal={Nature},
  volume={535},
  number={7610},
  pages={127--130},
  year={2016},
  publisher={Nature Publishing Group UK London}
}

@article{canales2023polaritonic,
  title={Polaritonic linewidth asymmetry in the strong and ultrastrong coupling regime},
  author={Canales, Adriana and Karmstrand, Therese and Baranov, Denis G and Antosiewicz, Tomasz J and Shegai, Timur O},
  journal={Nanophotonics},
  volume={12},
  number={21},
  pages={4073--4086},
  year={2023},
  publisher={De Gruyter}
}

@article{savona1995quantum,
  title={Quantum well excitons in semiconductor microcavities: Unified treatment of weak and strong coupling regimes},
  author={Savona, Vincenzo and Andreani, LC and Schwendimann, P and Quattropani, A},
  journal={Solid State Communications},
  volume={93},
  number={9},
  pages={733--739},
  year={1995},
  publisher={Elsevier}
}

@article{ebbesen2016hybrid,
  title={Hybrid light--matter states in a molecular and material science perspective},
  author={Ebbesen, Thomas W},
  journal={Accounts of chemical research},
  volume={49},
  number={11},
  pages={2403--2412},
  year={2016},
  publisher={ACS Publications}
}

@article{konrad2015controlling,
  title={Controlling the dynamics of F{\"o}rster resonance energy transfer inside a tunable sub-wavelength Fabry--P{\'e}rot-resonator},
  author={Konrad, Alexander and Metzger, Michael and Kern, Andreas M and Brecht, Marc and Meixner, Alfred J},
  journal={Nanoscale},
  volume={7},
  number={22},
  pages={10204--10209},
  year={2015},
  publisher={Royal Society of Chemistry}
}

@article{wang2016coherent,
  title={Coherent coupling of WS2 monolayers with metallic photonic nanostructures at room temperature},
  author={Wang, Shaojun and Li, Songlin and Chervy, Thibault and Shalabney, Atef and Azzini, Stefano and Orgiu, Emanuele and Hutchison, James A and Genet, Cyriaque and Samor{\`\i}, Paolo and Ebbesen, Thomas W},
  journal={Nano letters},
  volume={16},
  number={7},
  pages={4368--4374},
  year={2016},
  publisher={ACS Publications}
}

@article{kasprzak2006bose,
  title={Bose--Einstein condensation of exciton polaritons},
  author={Kasprzak, Jacek and Richard, Murielle and Kundermann, S and Baas, A and Jeambrun, P and Keeling, Jonathan Mark James and Marchetti, FM and Szyma{\'n}ska, MH and Andr{\'e}, R and Staehli, JL a and others},
  journal={Nature},
  volume={443},
  number={7110},
  pages={409--414},
  year={2006},
  publisher={Nature Publishing Group UK London}
}

@article{gu2021enhanced,
  title={Enhanced nonlinear interaction of polaritons via excitonic Rydberg states in monolayer WSe2},
  author={Gu, Jie and Walther, Valentin and Waldecker, Lutz and Rhodes, Daniel and Raja, Archana and Hone, James C and Heinz, Tony F and K{\'e}na-Cohen, St{\'e}phane and Pohl, Thomas and Menon, Vinod M},
  journal={Nature communications},
  volume={12},
  number={1},
  pages={2269},
  year={2021},
  publisher={Nature Publishing Group UK London}
}

@article{lidzey1999room,
  title={Room temperature polariton emission from strongly coupled organic semiconductor microcavities},
  author={Lidzey, DG and Bradley, DDC and Virgili, T and Armitage, A and Skolnick, MS and Walker, S},
  journal={Physical review letters},
  volume={82},
  number={16},
  pages={3316},
  year={1999},
  publisher={APS}
}

@article{lidzey1998strong,
  title={Strong exciton--photon coupling in an organic semiconductor microcavity},
  author={Lidzey, David G and Bradley, DDC and Skolnick, MS and Virgili, T and Walker, S and Whittaker, DM},
  journal={Nature},
  volume={395},
  number={6697},
  pages={53--55},
  year={1998},
  publisher={Nature Publishing Group UK London}
}

@article{hirai2024optical,
  title={Optical cavity design and functionality for molecular strong coupling},
  author={Hirai, Kenji and Andell Hutchison, James and Uji-i, Hiroshi},
  journal={Chemistry--A European Journal},
  volume={30},
  number={7},
  pages={e202303110},
  year={2024},
  publisher={Wiley Online Library}
}

@article{li2019tunable,
  title={Tunable open-access microcavities for solid-state quantum photonics and polaritonics},
  author={Li, Feng and Li, Yiming and Cai, Yin and Li, Peng and Tang, Haijun and Zhang, Yanpeng},
  journal={Advanced Quantum Technologies},
  volume={2},
  number={10},
  pages={1900060},
  year={2019},
  publisher={Wiley Online Library}
}

@article{hirai2023molecular,
  title={Molecular chemistry in cavity strong coupling},
  author={Hirai, Kenji and Hutchison, James A and Uji-i, Hiroshi},
  journal={Chemical Reviews},
  volume={123},
  number={13},
  pages={8099--8126},
  year={2023},
  publisher={ACS Publications}
}

@article{liu2015strong,
  title={Strong light--matter coupling in two-dimensional atomic crystals},
  author={Liu, Xiaoze and Galfsky, Tal and Sun, Zheng and Xia, Fengnian and Lin, Erh-chen and Lee, Yi-Hsien and K{\'e}na-Cohen, St{\'e}phane and Menon, Vinod M},
  journal={Nature Photonics},
  volume={9},
  number={1},
  pages={30--34},
  year={2015},
  publisher={Nature Publishing Group UK London}
}

@article{chen20242d,
  title={A 2D chiral microcavity based on apparent circular dichroism},
  author={Chen, Tzu-Ling and Salij, Andrew and Parrish, Katherine A and Rasch, Julia K and Zinna, Francesco and Brown, Paige J and Pescitelli, Gennaro and Urraci, Francesco and Aronica, Laura A and Dhavamani, Abitha and others},
  journal={Nature communications},
  volume={15},
  number={1},
  pages={3072},
  year={2024},
  publisher={Nature Publishing Group UK London}
}

@article{koks2021microcavity,
  title={Microcavity resonance condition, quality factor, and mode volume are determined by different penetration depths},
  author={Koks, Corn{\'e} and Van Exter, MP},
  journal={Optics Express},
  volume={29},
  number={5},
  pages={6879--6889},
  year={2021},
  publisher={Optical Society of America}
}

\bibliographyfullrefs{sample}


\ifthenelse{\equal{\journalref}{aop}}{%
\section*{Author Biographies}
\begingroup
\setlength\intextsep{0pt}
\begin{minipage}[t][6.3cm][t]{1.0\textwidth} 
  \begin{wrapfigure}{L}{0.25\textwidth}
    \includegraphics[width=0.25\textwidth]{john_smith.eps}
  \end{wrapfigure}
  \noindent
  {\bfseries John Smith} received his BSc (Mathematics) in 2000 from The University of Maryland. His research interests include lasers and optics.
\end{minipage}
\begin{minipage}{1.0\textwidth}
  \begin{wrapfigure}{L}{0.25\textwidth}
    \includegraphics[width=0.25\textwidth]{alice_smith.eps}
  \end{wrapfigure}
  \noindent
  {\bfseries Alice Smith} also received her BSc (Mathematics) in 2000 from The University of Maryland. Her research interests also include lasers and optics.
\end{minipage}
\endgroup
}{}

\end{document}